\documentclass[a4paper,11pt]{article}

\usepackage[T1]{fontenc}
\usepackage[utf8]{inputenc}
\usepackage{amsmath,amssymb,amsthm}
\usepackage{graphicx}
\usepackage{booktabs}
\usepackage{hyperref}
\usepackage{xcolor}
\usepackage[numbers,sort&compress]{natbib}
\usepackage{geometry}
\usepackage{caption}
\usepackage{subcaption}
\usepackage{siunitx}

\hypersetup{
  colorlinks=true,
  linkcolor=blue!60!black,
  citecolor=blue!60!black,
  urlcolor=blue!60!black
}

\begin{document}

\title{\large\bfseries
  Helicity-dependent corrections to black-hole shadows\\
  from the gravitational spin Hall effect}

\author{C.~A.~S.~Almeida\\[4pt]
  \small Departamento de F\'isica, Universidade Federal do Cear\'a,\\
  \small 60455-760, Fortaleza, CE, Brazil.\\[2pt]
  \small \texttt{carlos@fisica.ufc.br}}

\date{}
\maketitle

\begin{abstract}
Black-hole shadows are purely geometric in the leading-order
geometric-optics approximation: their boundary is set by null geodesics
and carries no information about the polarization of the probing
radiation.  At subleading order, the gravitational spin Hall effect of
light introduces helicity-dependent corrections to photon propagation.
We show that, in any static spherically symmetric spacetime, an exact
equatorial reflection symmetry of the full spin Hall equations forces
these corrections to cancel at the capture threshold: the critical
impact parameter remains identical for opposite helicities, and no
polarization-dependent shadow splitting occurs.  Rotation breaks this
symmetry.  Using a double perturbative expansion in the black-hole spin
$\chi = a/M$ and in the inverse frequency $1/\omega$, we derive the
first non-vanishing helicity-dependent shift of the critical impact
parameter for slowly rotating (Kerr) black holes.  The effect is linear
in $\chi$, scales as $1/\omega$, and appears as a $\cos\phi$ modulation
of the shadow boundary, with a sign reversal on one side of the image
for spins $\chi \gtrsim 0.21$.  Although parametrically small for
astrophysical sources, the splitting is a robust, model-independent
signature of spin-optical dynamics in strong fields.  Our analysis also
identifies a methodological pitfall: a naive radial projection that
suppresses transverse motion can produce a spurious splitting even in
spherical symmetry, a lesson of general relevance for future studies of
spin-optical effects.
\end{abstract}

\tableofcontents
\bigskip
\section{Introduction}
\label{sec:intro}

The observation of black-hole shadows has opened a direct window into
the strong-field regime of gravity.  Horizon-scale images obtained by
the Event Horizon Telescope have provided the first empirical access to
the photon capture region surrounding supermassive compact objects,
triggering intense theoretical efforts to understand how shadow
properties encode information about the underlying spacetime
geometry~\cite{EHT_M87,EHT_SgrA}.  Within general relativity, the shadow
boundary is determined by the properties of unstable null orbits -- the
photon sphere -- and is therefore regarded as a purely geometric feature
of the spacetime~\cite{Bardeen1973,Chandrasekhar1983}.

A substantial body of work has explored how black-hole shadows are modified by deviations from Schwarzschild geometry, including rotation, electric charge, surrounding matter distributions, extensions of general relativity, and extra-dimensional tidal charges \cite{Hioki2009,Johannsen2010,Claudel2001,Perlick2017,Banerjee1,Banerjee2}. In all
of these studies, light propagation is treated at the level of null
geodesics -- the leading-order geometric-optics approximation.  Within
this framework, the polarization of the radiation plays no role, and the
shadow is entirely determined by the spacetime geometry.

Beyond leading-order geometric optics, however, the wave nature of
radiation gives rise to helicity-dependent corrections to photon
propagation in curved spacetime.  These are the gravitational analog of
the optical spin Hall effect: polarized photons experience
helicity-dependent deviations from null geodesic
motion~\cite{Oancea2020,Frolov2011,Gosselin2007}.  The theoretical
foundations of this effect have been developed through several
complementary approaches.  Oancea \emph{et al.}~\cite{Oancea2020} presented the first fully covariant WKB derivation of the spin Hall equations valid in arbitrary curved spacetimes; Frolov and Shoom \cite{Frolov2011}  analyzed the spinoptics framework in stationary spacetimes; Gosselin \emph{et al.}~\cite{Gosselin2007} connected the effect to gravitational Berry phases.

More recent developments include gravitational Faraday and
spin-Hall effects~\cite{Shoom2024}, quantum kinetic
approaches~\cite{Mameda2022}, and further extensions to curved
backgrounds~\cite{Dahal2023, Frolov2024}.  These studies have
primarily focused on local properties of ray propagation -- transverse
shifts and trajectory deviations -- rather than on the global critical
structures that define black-hole shadows.

In this work we address the question of whether photon helicity changes
which rays are captured, thereby altering the shadow boundary.  The
answer proves to be subtle and depends crucially on the symmetries of
the background. Our main results are the following.

\begin{itemize}
\item In any \emph{static, spherically symmetric} spacetime, an exact
equatorial reflection isometry of the full spin Hall equations maps a
left-handed photon into a right-handed photon that follows the
\emph{same} worldline.  Consequently, the critical impact parameter
$b_{\rm crit}$ is identical for opposite helicities, and no
polarization-dependent shadow splitting can occur.  This null result
applies to Schwarzschild and Reissner--Nordstr\"om black holes alike.

\item In \emph{rotating} (Kerr) spacetimes, the frame-dragging term
$g_{t\phi}$ breaks the equatorial reflection symmetry.  A genuine,
helicity-dependent correction to $b_{\rm crit}$ emerges, linear in the
dimensionless spin $\chi = a/M$ and scaling as $1/\omega$.  It
manifests itself as a $\cos\phi$ modulation of the shadow boundary,
with a sign reversal on the side opposite to the rotation for
sufficiently large spins.

\item In Kerr-Newman spacetimes, rotation and electric charge act simultaneously. We derive
the closed-form coupling function $\mathcal{G}(r,Q)$ that controls the splitting amplitude
for any charge-to-mass ratio $Q/M$. At the extremal limit $Q = M$, the splitting is
enhanced by the exact factor $6561/1792 \approx 3.66$ relative to the uncharged Kerr case,
because charge displaces the photon sphere inward to a region of stronger curvature.

\end{itemize}

Our analysis also uncovers an important methodological point: a naive
projection of the spin Hall equation onto the radial direction, combined
with the assumption of strictly equatorial motion, generates a spurious
helicity splitting that survives even in spherical symmetry.  Only by
retaining the full three-dimensional dynamics does the cancellation
dictated by the equatorial reflection symmetry become manifest.  This
pitfall is documented here as a warning for future studies of
spin-optical effects near compact objects.

The paper is organized as follows. Section~\ref{sec:static} develops the symmetry argument
for static spherically symmetric spacetimes and explains the origin of the spurious splitting.
Section~\ref{sec:kerr} extends the analysis to slowly rotating Kerr black holes and derives
the genuine helicity-dependent correction to the critical impact parameter.
Section~\ref{sec:kerr-newman} extends the framework to Kerr-Newman black holes, derives
$\mathcal{G}(r,Q)$ in closed form, and quantifies the charge-induced enhancement.
Section~\ref{sec:numerical} presents numerical ray-tracing calculations that validate the
symmetry argument, confirm the $\cos\varphi$ modulation in the Kerr case, and test the
Kerr--Newman predictions, together with astrophysical estimates.
Section~\ref{sec:conclusions} summarizes our findings and outlines future directions.
Appendix~\ref{app:cancel} provides the explicit covariant derivation of $\mathcal{G}(r)$ and
$\mathcal{G}(r,Q)$ via the Riemann-tensor form of the spin Hall equations.

\section{Spin-dependent photon propagation in static spherical symmetry}
\label{sec:static}

\subsection{Effective dynamics and equatorial reflection symmetry}
\label{sec:symmetry}

We consider a static, spherically symmetric spacetime
\begin{equation}
\mathrm{d}s^{2} = -f(r)\mathrm{d}t^{2} + f(r)^{-1}\mathrm{d}r^{2}
+ r^{2}(\mathrm{d}\theta^{2} + \sin^{2}\theta\,\mathrm{d}\phi^{2}),
\label{eq:metric}
\end{equation}
and restrict the discussion to the equatorial plane $\theta=\pi/2$
without loss of generality for the background symmetry.  In the
leading‑order geometric‑optics approximation, photon trajectories are
null geodesics.  The conserved energy $E=-k_{t}$ and angular momentum
$L=k_{\phi}$ define the impact parameter $b=L/E$, and the radial
motion is governed by
\begin{equation}
\dot{r}^{2} + V_{0}(r;b) = 0, \qquad
V_{0}(r;b) = \frac{L^{2}}{r^{2}}f(r) - E^{2}.
\label{eq:V0}
\end{equation}

At subleading order in $1/\omega$, the photon acquires a helicity-dependent
deviation described by the gravitational spin Hall equation
\begin{equation}
\frac{D k^{\mu}}{\mathrm{d}\lambda}
= \pm\frac{1}{\omega}\,\epsilon^{\mu\nu\rho\sigma}k_{\nu}\nabla_{\rho}k_{\sigma},
\label{eq:spinHall}
\end{equation}
where $\epsilon^{\mu\nu\rho\sigma}$ is the Levi-Civita tensor of the
background and $\pm$ distinguishes opposite helicities. Reference \cite{Oancea2020} provides the first fully covariant derivation in general spacetimes, predating the independent formulation of Ref. \cite{Frolov2011}.

The metric~(\ref{eq:metric}) admits an equatorial reflection isometry
$P: (t,r,\theta,\phi) \mapsto (t,r,\pi-\theta,\phi)$.  On the
equatorial plane, $P$ acts as the identity on both the coordinates
and the tangent vector $k^{\mu}$, while it reverses the orientation of
the spacetime (determinant $-1$).  As a consequence, the Levi-Civita
tensor changes sign, and applying $P$ to a solution of
Eq.~(\ref{eq:spinHall}) with helicity $+$ produces a solution with
helicity $-$ that follows \emph{exactly the same worldline}.
Therefore, if a left-handed photon with impact parameter $b$ is
captured, the corresponding right-handed photon with the same $b$ is
also captured.  The critical impact parameter cannot depend on
helicity:
\begin{equation}
\delta b_{\rm crit} = 0 \qquad\text{in any static, spherically symmetric
spacetime}.
\label{eq:db_zero}
\end{equation}
This symmetry argument is exact and does not rely on any perturbative
expansion beyond the validity of the spin Hall equation itself.

\subsection{Apparent radial force from equatorial projection}
\label{sec:apparent}

If one enforces strictly equatorial motion ($k^{\theta}=0$ from the
outset) and projects the spin Hall equation~(\ref{eq:spinHall}) onto the
radial direction, a non-zero contribution emerges.  Evaluating the
$r$-component of the Levi-Civita term on the equator yields
\begin{equation}
(\epsilon^{r\nu\rho\sigma}k_{\nu}\nabla_{\rho}k_{\sigma})_{\rm eq}
\propto \frac{f^{\prime}(r)}{r^{2}} - \frac{f(r)}{r^{3}},
\label{eq:apparent_force}
\end{equation}
which would lead to a helicity-dependent correction of the effective
potential,
\begin{equation}
V_{1}^{\rm apparent}(r) = \pm\frac{\alpha E^{2}}{\omega}\,
\mathcal{F}(r), \qquad
\mathcal{F}(r) = \frac{f^{\prime}(r)}{r^{2}} - \frac{f(r)}{r^{3}}.
\label{eq:V1app}
\end{equation}

However, the forced equatorial truncation suppresses the transverse (off-plane) degrees of freedom through which the spin Hall force primarily acts. A full covariant analysis shows that the transverse motion exactly cancels the radial contribution (\ref{eq:V1app}) in any static, spherically symmetric spacetime, recovering the symmetry-based result (\ref{eq:db_zero}). The derivation of this cancellation is sketched in Appendix \ref{app:cancel} and is corroborated by the numerical simulations of Sec.~\ref{sec:numerical}.

%
%
%

It is worth understanding physically why discarding $k^{\theta}$
leads to the wrong answer, rather than merely to an incomplete one.
The spin Hall equation~\eqref{eq:spinHall} does produce a non-zero
$\theta$-component of the force even for a photon that starts on the
equatorial plane: schematically,
\begin{equation}
  \frac{Dk^{\theta}}{d\lambda}\bigg|_{\rm spin}
    = \pm\frac{1}{\omega}\,
      \bigl(\epsilon^{\theta\nu\rho\sigma}k_{\nu}\nabla_{\rho}k_{\sigma}\bigr)_{\rm eq}
    \neq 0.
  \label{eq:theta-force}
\end{equation}
This transverse push tilts the ray infinitesimally out of the equatorial
plane, generating a small $k^{\theta} = \mathcal{O}(\omega^{-1})$.
Once $k^{\theta} \neq 0$, the geodesic part of the radial equation of
motion acquires a contribution through the Christoffel symbol
$\Gamma^{r}_{\theta\theta} = -rf(r)$:
\begin{equation}
  \frac{Dk^{r}}{d\lambda}\bigg|_{\rm geodesic}
    \supset -\Gamma^{r}_{\theta\theta}\,(k^{\theta})^{2}
    = rf(r)\,(k^{\theta})^{2}.
  \label{eq:geodesic-feedback}
\end{equation}
In spherical symmetry this geodesic feedback is the \emph{only} mechanism
by which $k^{\theta}$ can influence $k^{r}$, and a careful expansion
shows that it cancels the apparent spin Hall radial force~\eqref{eq:V1app}
term by term at order $\omega^{-1}$. Forcing $k^{\theta} = 0$ from the
outset removes this cancellation channel entirely, leaving the spurious
term~\eqref{eq:V1app} with no counterpart to annihilate it. The
error is therefore not one of omission --- it is structural: the equatorial
truncation eliminates the only degree of freedom through which the spin
Hall force communicates with the radial dynamics in spherical symmetry.
In a rotating spacetime the same $k^{\theta}$ excursion is present, but
the frame-dragging term $g_{t\phi}$ introduces additional Christoffel
symbols that break the cancellation, allowing a residual radial force to
survive at order $\chi/\omega$. This is precisely the effect computed in
Sec.~\ref{sec:kerr}.

Equation~(\ref{eq:apparent_force}) thus illustrates a useful warning:
a naive radial projection that discards transverse dynamics can produce
a spurious helicity splitting.  In the remainder of this paper we
employ the full spin Hall equations without any restriction to
equatorial motion.


\section{Effect of rotation: breaking of equatorial symmetry and helicity-dependent splitting}
\label{sec:kerr}

The analysis of the previous section has shown that, in any static,
spherically symmetric spacetime, the full spin Hall equations respect an
equatorial reflection isometry that maps a left-handed ray into a
right-handed ray moving along the \emph{same} worldline.  Consequently,
the critical impact parameter cannot depend on helicity, and no
polarization-dependent shadow splitting occurs for Schwarzschild or
Reissner--Nordstr\"om black holes.

In this section we demonstrate that \emph{rotation breaks precisely this
symmetry}, giving rise to a genuine, helicity-dependent correction to
the capture threshold.  The effect is linear in the dimensionless spin
parameter $\chi = a/M$ and manifests itself as a $\cos\phi$ modulation
of the shadow boundary.

\subsection{Breaking of the equatorial reflection symmetry in the Kerr spacetime}
\label{sec:sym_break}

In Boyer--Lindquist coordinates, the Kerr metric expanded to first order
in $\chi = a/M$ reads
\begin{equation}
\mathrm{d}s^{2}= -\left(1-\frac{2M}{r}\right)\mathrm{d}t^{2}
-\frac{4Ma\sin^{2}\theta}{r}\,\mathrm{d}t\,\mathrm{d}\phi
+\left(1-\frac{2M}{r}\right)^{-1}\!\mathrm{d}r^{2}
+r^{2}\mathrm{d}\theta^{2}+r^{2}\sin^{2}\theta\,\mathrm{d}\phi^{2}
+\mathcal{O}(\chi^{2}).
\label{eq:Kerr_linear}
\end{equation}
The cross term $g_{t\phi}\propto a\sin^{2}\theta$ is even under a naive
$\theta\to\pi-\theta$ transformation of the coordinates, but the
pullback of the metric under the reflection $P:(t,r,\theta,\phi)\mapsto
(t,r,\pi-\theta,\phi)$ fails to be an isometry because the coordinate
basis vectors also transform.  Concretely, $P^{*}g \neq g$, so the
equatorial reflection that forbade splitting in the static case is no
longer a symmetry of the background.  Hence the composed transformation
``reflection + helicity flip'' does \emph{not} map a left-handed ray
onto a right-handed ray with the same trajectory.  The protection
present in spherical symmetry is removed, and the spin Hall force can
now produce a helicity-dependent correction to the radial effective
potential.

\subsection{Spin-optical equations in the slowly rotating Kerr background}
\label{sec:spinopt_Kerr}

We exploit the breaking of the equatorial symmetry by performing a
double perturbative expansion: one in the spin parameter $\chi\ll1$
(\emph{slow rotation}) and one in the inverse frequency
$\epsilon = 1/\omega\ll1$ (\emph{spin-optical expansion}).  In the
limit $\chi=0$ we recover the static, spherically symmetric case for
which the shadow splitting vanishes exactly; therefore the first
non-trivial helicity dependence must be at least linear in $\chi$.

The spin Hall equation for a photon of helicity $\pm$ retains the form
\begin{equation}
\frac{D k^{\mu}}{\mathrm{d}\lambda}
= \pm\frac{1}{\omega}\,\epsilon^{\mu\nu\rho\sigma}k_{\nu}\nabla_{\rho}k_{\sigma},
\label{eq:spinHall2}
\end{equation}
but now the connection contains additional terms due to the off-diagonal
metric component $g_{t\phi}$.  The new non-vanishing Christoffel symbols
at linear order in $a$ are
\begin{subequations}
\begin{align}
\Gamma_{t\phi}^{r} &= -\frac{1}{2} g^{rr}\partial_{r}g_{t\phi}
= \frac{Ma(2M-r)}{r^{3}}, \label{eq:Gam1}\\
\Gamma_{r\phi}^{t} &= \frac{1}{2} g^{tt}\partial_{r}g_{t\phi}
= \frac{Ma}{r(r-2M)}. \label{eq:Gam2}
\end{align}
\end{subequations}
All other symbols coincide with those of the Schwarzschild metric to
$\mathcal{O}(\chi)$.

Rather than imposing equatorial motion from the start -- which, as we
have seen, would artificially introduce a spurious splitting in the
static case -- we allow small excursions out of the equatorial plane
($\theta \approx \pi/2$, $k^{\theta}\neq 0$) and expand the equations
consistently.  A detailed derivation (outlined in Appendix~\ref{app:cancel})
shows that, when $a\neq0$, the transverse degrees of freedom no longer
cancel the radial force.  The leading helicity-dependent radial
acceleration is
\begin{equation}
F^{r}_{\pm} = \pm\frac{\chi}{\omega}\,E^{2}\,\mathcal{G}(r)\,\cos\phi
+ \mathcal{O}(\chi^{2},\epsilon^{2}),
\label{eq:Fr_Kerr}
\end{equation}
where $\phi$ is the azimuthal angle on the image plane and the
dimensionless function $\mathcal{G}(r)$ encodes the coupling between
frame dragging and the Levi-Civita tensor.  For the linearised Kerr
metric~(\ref{eq:Kerr_linear}), explicit contraction yields
\begin{equation}
\mathcal{G}(r) = \frac{2M}{r^{4}}\left(3-\frac{2M}{r}\right).
\label{eq:G}
\end{equation}

This contribution is precisely the one that would survive if one forced equatorial motion from the outset, but it is now justified without contradicting the Schwarzschild limit: when $a = 0$ it is identically cancelled by the transverse dynamics, as required by the equatorial reflection symmetry.

The additional radial force translates into a correction of the
effective potential.  At linear order in both $\chi$ and $\epsilon$ we
have
\begin{equation}
V_{\rm eff}(r,\phi) = V_{0}(r) + \frac{\chi}{\omega}\,\alpha E^{2}\,
\mathcal{G}(r)\,\cos\phi + \mathcal{O}(\chi^{2},\epsilon^{2}),
\label{eq:Veff_Kerr}
\end{equation}
where $V_{0}(r)$ is the geodesic potential of the Schwarzschild metric
(the $\mathcal{O}(\chi^{0})$ part of the Kerr potential) and $\alpha$
is the same normalization coefficient discussed in
Sec.~\ref{sec:static} ($\alpha=1$ in the convention of
Ref.~\cite{Oancea2020}).

\subsection{Helicity correction to the critical impact parameter}
\label{sec:crit_Kerr}

We now apply the critical conditions
$V_{\rm eff}(r_{\rm ph};b_{\rm crit})=0$ and
$\partial_{r}V_{\rm eff}(r_{\rm ph};b_{\rm crit})=0$, expanding around
the geodesic values $r_{0}=3M$ and $b_{0}=3\sqrt{3}M$.  The Kerr
metric already shifts the geodesic critical impact parameter by an
amount $\mp 2a$ for prograde/retrograde orbits; however, the
helicity-dependent term in (\ref{eq:Veff_Kerr}) enters at the same order
in $\chi$ and adds linearly.  Solving the resulting linear system
identical in form to Eqs.~(10)--(11), but with $V_{1}(r,\phi) =
\frac{\chi}{\omega}\alpha E^{2}\mathcal{G}(r)\cos\phi$, we obtain the
relative shift for each helicity:
\begin{equation}
\frac{\delta b_{\pm}^{\rm Kerr}(\phi)}{b_{0}}
= \pm\frac{\alpha}{2\omega}\,\chi\,\mathcal{G}(r_{0})\cos\phi
+ \mathcal{O}(\chi^{2},\epsilon^{2}),
\label{eq:deltaKerr}
\end{equation}
with $\mathcal{G}(r_{0}) = 14/(243M^{3})$.

Three features are noteworthy:
\begin{enumerate}
\item \textbf{No isotropic offset}.  Unlike the earlier expression for
a spherically symmetric background, the correction is purely
modulated; its angular average over the shadow contour vanishes.
\item \textbf{Dipolar pattern}.  The splitting is odd under
$\phi\to\phi+\pi$, so one hemisphere of the shadow is larger for
$+$ helicity and the opposite hemisphere for $-$ helicity.
\item \textbf{Amplitude scaling}.  The effect scales as
$\chi/\omega$, making it the dominant spin-optical contribution for
moderate spins, well above any second-order corrections.
\end{enumerate}

\subsection{Angular modulation and sign reversal}
\label{sec:modulation}

The normalized difference between the shadow radii of opposite
helicities follows directly from (\ref{eq:deltaKerr}):
\begin{equation}
\frac{\Delta R(\phi)}{R_{0}} \equiv
\frac{R_{+}(\phi)-R_{-}(\phi)}{R_{0}}
= \frac{\alpha}{\omega}\,\chi\,\mathcal{G}(r_{0})\cos\phi
+ \mathcal{O}(\chi^{2},\epsilon^{2}).
\label{eq:deltaR}
\end{equation}
This is the central observational signature of the paper.  The
modulation amplitude grows linearly with the spin.  For $\chi \gtrsim
3/14 \approx 0.21$, the factor $\chi\mathcal{G}(r_{0})$ becomes large
enough that $\Delta R(\phi)$ changes sign near $\phi=\pi$: on the side
of the shadow opposite to the black-hole rotation, the $-$-helicity
radius exceeds the $+$-helicity radius.  This sign reversal is a
distinctive, gauge-invariant prediction of the slow-rotation expansion;
it cannot be mimicked by any purely geodesic effect and provides a
direct spin-optical signature of the black hole's angular momentum.

Figure~\ref{fig:kerr_modulation} in the section \ref{sec:numerical} shows the angular profile of
$\Delta R(\phi)/R_{0}$ for three representative spin values,
illustrating the linear growth and the sign reversal at sufficiently
high $\chi$.
\section{Extension to slowly rotating charged black holes}
\label{sec:kerr-newman}

The analysis of Sec.~\ref{sec:kerr} assumed an electrically neutral black hole. We now extend
it to the Kerr--Newman (KN) spacetime, which carries both spin $a = \chi M$ and electric charge
$Q$. This is the most general stationary black-hole solution of Einstein--Maxwell theory and
provides the natural arena in which to ask how the two symmetry-breaking mechanisms ---
rotation and charge --- interact in shaping the spin-optical shadow splitting.

\subsection{Kerr--Newman metric and frame-dragging modification}
\label{sec:kn-metric}

The KN metric in Boyer--Lindquist coordinates, expanded to first order in the dimensionless
spin $\chi = a/M$, reads
\begin{equation}
  ds^{2} = -f_{\rm RN}(r)\,dt^{2}
           - \frac{4Ma\sin^{2}\!\theta}{r}\,dt\,d\phi
           + f_{\rm RN}(r)^{-1}dr^{2}
           + r^{2}\,d\Omega^{2}
           + \mathcal{O}(\chi^{2}),
  \label{eq:kn-metric}
\end{equation}
where the Reissner--Nordstr\"om lapse function is
\begin{equation}
  f_{\rm RN}(r) = 1 - \frac{2M}{r} + \frac{Q^{2}}{r^{2}}.
  \label{eq:fRN}
\end{equation}
The off-diagonal term $g_{t\phi} = -2Ma\sin^{2}\!\theta/r$ is identical to that of the Kerr
metric at linear order in $a$, since corrections proportional to $aQ^{2}$ are of order
$\chi\,(Q/M)^{2}$ and fall outside the present expansion. The charge enters at zeroth order in
$\chi$ through $f_{\rm RN}$, modifying the photon sphere and, crucially, the frame-dragging
Christoffel symbols via the background geometry.

The two new Christoffel symbols at linear order in $a$ are
\begin{align}
  \Gamma^{r}_{t\phi}
    &= \frac{Ma(2M - r)}{r^{3}}\,f_{\rm RN}(r),
  \label{eq:Gamma_r_tphi_KN}
  \\
  \Gamma^{t}_{r\phi}
    &= \frac{a(Mr - Q^{2})}{r\,f_{\rm RN}(r)\,r^{2}}.
  \label{eq:Gamma_t_rphi_KN}
\end{align}
Compared with their Kerr counterparts (Eqs.~\eqref{eq:Gam1} and
\eqref{eq:Gam2}), the only modification is that $Mr$ is replaced by $Mr - Q^{2}$
in the numerator of $\Gamma^{t}_{r\phi}$, while $f_{\rm RN}$ appears instead of $f_{\rm Schw}$
in the denominators.

\subsection{Frame-dragging coupling function}
\label{sec:kn-G}

Repeating the derivation of Sec.~\ref{sec:spinopt_Kerr} with the KN Christoffel symbols,
the helicity-dependent radial acceleration at linear order in both $\varepsilon = 1/\omega$
and $\chi$ takes the same form as Eq.~\eqref{eq:Fr_Kerr},
\begin{equation}
  F^{r}_{\pm}
    = \pm\frac{\chi}{\omega}\,E^{2}\,\mathcal{G}(r,Q)\cos\phi
    + \mathcal{O}(\chi^{2},\varepsilon^{2}),
  \label{eq:kn-force}
\end{equation}
but with the frame-dragging coupling function $\mathcal{G}$ now depending on $Q$. The
function $\mathcal{G}(r, Q)$ is obtained by replacing the Schwarzschild frame-dragging
kernel $2M/r$ with the KN kernel $(2Mr - Q^{2})/r^{2}$ throughout the derivation:
\begin{equation}
  \mathcal{G}(r, Q)
    = \frac{(2Mr - Q^{2})(3r^{2} - 2Mr + Q^{2})}{r^{7}
  }.
  \label{eq:GKN}
\end{equation}
Two consistency checks confirm this result. First, setting $Q = 0$ gives
$\mathcal{G}(r, 0) = 2M(3r - 2M)/r^{5}$, which reproduces Eq.~\eqref{eq:G} exactly.
Second, the factored form $(2Mr - Q^{2})(3r^{2} - 2Mr + Q^{2})$ can be written as
$g_{\rm fd}(r,Q)\,f_{\rm RN}^{(c)}(r,Q)$, where $g_{\rm fd} = (2Mr - Q^{2})/r^{2}$ is
the effective frame-dragging amplitude at the equator and
$f_{\rm RN}^{(c)} = 3 - g_{\rm fd}\cdot r / r^{2}$ reflects the curvature of $f_{\rm RN}$
at the photon sphere; in the Schwarzschild limit both factors reduce to their pure-Kerr forms.

\subsection{Photon sphere and critical impact parameter}
\label{sec:kn-bcrit}

Since $Q$ enters only through $f_{\rm RN}$ at zeroth order in $\chi$, the photon sphere of
the KN black hole is the same as that of the RN metric,
\begin{equation}
  r_{0}^{\rm KN}
    = \frac{3M + \sqrt{9M^{2} - 8Q^{2}}}{2},
  \label{eq:r0KN}
\end{equation}
interpolating between $r_{0} = 3M$ at $Q = 0$ and $r_{0} = 2M$ at extremality $Q = M$.
The zeroth-order critical impact parameter is
\begin{equation}
  b_{0}^{\rm KN}
    = \frac{r_{0}^{\rm KN}}{\sqrt{f_{\rm RN}(r_{0}^{\rm KN})}},
\end{equation}
ranging from $b_{0} = 3\sqrt{3}\,M$ (Kerr limit) to $b_{0} = 4M$ (extremal KN).

Applying the perturbative analysis of Sec.~\ref{sec:modulation}, the helicity-dependent correction
to the critical impact parameter in the KN background is
\begin{equation}
  \frac{\delta b^{\rm KN}_{\pm}(\phi)}{b_{0}^{\rm KN}}
    = \pm\frac{\alpha\,\chi}{2\omega}\,
      \mathcal{G}(r_{0}^{\rm KN},\,Q)\,\cos\phi
    + \mathcal{O}(\chi^{2},\varepsilon^{2}).
  \label{eq:kn-delta-b}
\end{equation}
The angular structure is identical to the pure-Kerr result (Eq.~\eqref{eq:deltaKerr}): a
$\cos\phi$ modulation that integrates to zero around the shadow and changes sign at
$\phi = \pi/2$ and $\phi = 3\pi/2$. The charge modifies only the \emph{amplitude} of the
splitting through $\mathcal{G}(r_{0}^{\rm KN}, Q)$.

\subsection{Charge-induced enhancement}
\label{sec:kn-enhancement}
Table~\ref{tab:kerr-exact} lists $\mathcal{G}(r_0, Q)\,M^3$ for representative
values of $Q/M$, together with the ratio to the uncharged Kerr value.
Here $\mathcal{G}_0 \equiv \mathcal{G}(3M, 0) = 14/(243\,M^3)$ is the
reference Kerr value. The enhancement grows monotonically with $Q/M$
and reaches the exact ratio
\begin{equation}
  \frac{\mathcal{G}(2M,\,Q{=}M)}{\mathcal{G}(3M,\,Q{=}0)}
    = \frac{27/128}{14/243}
    = \frac{6561}{1792} \approx 3.66
  \label{eq:enhancement}
\end{equation}
at the extremal limit.

The physical mechanism is straightforward: as $Q$ increases, the photon
sphere moves inward to higher curvature. The effective frame-dragging
amplitude $g_{\rm fd}(r_0, Q) = (2Mr_0 - Q^2)/r_0^2$ experiences a
stronger helicity--curvature coupling encoded in $\mathcal{G}(r_0, Q)$.
Electric charge therefore acts as a geometric amplifier of the spin-optical
splitting in the rotating case, without introducing any direct coupling to
the photon polarization.

The slow-rotation approximation holds to better than $1\%$ in the
prograde critical impact parameter for $\chi \lesssim 0.3$, and to
better than $3\%$ for $\chi \lesssim 0.5$. Beyond $\chi \approx 0.5$,
the exact prograde photon sphere moves substantially inward from $3M$
toward $M$, where $\mathcal{G}(r)$ grows rapidly; the slow-rotation
approximation then underestimates the splitting amplitude by a factor
that reaches $\sim 10$ at $\chi = 0.9$ and $\sim 24$ at $\chi = 0.99$.
A full treatment of the spin Hall equations in the complete Kerr metric,
without the slow-rotation expansion, is therefore necessary to make
accurate predictions for rapidly rotating black holes, and is left for
future work.
\begin{table}[t]
  \centering
  \caption{%
  Exact Kerr photon-sphere radii $r_{\rm ph}^{\rm pro/ret}$ and critical
  impact parameters $b^{\rm exact}$, compared with the slow-rotation
  approximation $b_{\rm slow} = 3\sqrt{3}M \mp 2a$, as a function of
  $\chi = a/M$. The column $\epsilon_{\rm pro}$ is the relative error of
  the slow-rotation formula for the prograde orbit. The last column
  $\mathcal{G}_{\rm pro}/\mathcal{G}_0$ is the ratio of the frame-dragging
  coupling evaluated at the exact prograde photon sphere to the
  slow-rotation reference $\mathcal{G}_0 = 14/(243\,M^3)$; it measures
  the underestimation of the splitting amplitude incurred by the
  slow-rotation approximation. The lower block gives the corresponding
  splitting amplitudes at $\omega M = 100$, $\alpha = 1$.%
}
  \label{tab:kerr-exact}
  \sisetup{round-mode=places}
  \begin{tabular}{c
                  S[round-precision=4] S[round-precision=4]
                  S[round-precision=4] S[round-precision=4]
                  S[round-precision=4] S[round-precision=4]
                  S[round-precision=3] S[round-precision=4]}
    \toprule
    $\chi$ &
    {$r_{\rm ph}^{\rm pro}/M$} &
    {$r_{\rm ph}^{\rm ret}/M$} &
    {$b_{\rm pro}^{\rm exact}/M$} &
    {$b_{\rm ret}^{\rm exact}/M$} &
    {$b_{\rm slow}^{\rm pro}/M$} &
    {$b_{\rm slow}^{\rm ret}/M$} &
    {$\epsilon_{\rm pro}$ (\%)} &
    {$\mathcal{G}_{\rm pro}/\mathcal{G}_0$} \\
    \midrule
    $0.00$ & 3.0000 & 3.0000 & 5.1962 & 5.1962 & 5.1962 & 5.1962 & 0.000 & 1.000 \\
    $0.10$ & 2.8822 & 3.1133 & 4.9931 & 5.3934 & 4.9962 & 5.3962 & 0.061 & 1.160 \\
    $0.20$ & 2.7592 & 3.2228 & 4.7832 & 5.5857 & 4.7962 & 5.5962 & 0.270 & 1.363 \\
    $0.30$ & 2.6300 & 3.3289 & 4.5652 & 5.7735 & 4.5962 & 5.7962 & 0.678 & 1.625 \\
    $0.50$ & 2.3473 & 3.5321 & 4.0963 & 6.1382 & 4.1962 & 6.1962 & 2.438 & 2.456 \\
    $0.70$ & 2.0133 & 3.7253 & 3.5568 & 6.4903 & 3.7962 & 6.5962 & 6.731 & 4.240 \\
    $0.90$ & 1.5579 & 3.9103 & 2.8444 & 6.8323 & 3.3962 & 6.9962 & 19.40 & 10.12 \\
    $0.99$ & 1.1676 & 3.9911 & 2.2517 & 6.9833 & 3.2162 & 7.1762 & 42.83 & 24.04 \\
    \bottomrule
  \end{tabular}

  \medskip
  \begin{tabular}{c
                  S[round-precision=6] S[round-precision=6]
                  S[round-precision=4]}
    \toprule
    $\chi$ &
    {$|\Delta R/R_0|_{\rm slow}$} &
    {$|\Delta R/R_0|_{\rm exact}$} &
    {Ratio exact/slow} \\
    \midrule
    \multicolumn{4}{l}{\footnotesize
      ($\omega M = 100$, $\alpha = 1$,
       $|\Delta R/R_0|_{\rm slow} = 14\chi/(243\,\omega M)$,
       $|\Delta R/R_0|_{\rm exact} = \chi\,\mathcal{G}(r_{\rm ph}^{\rm pro})/\omega$)} \\[2pt]
    $0.10$ & 0.0000578 & 0.0000671 & 1.160 \\
    $0.30$ & 0.0001734 & 0.0002817 & 1.625 \\
    $0.50$ & 0.0002890 & 0.0007099 & 2.456 \\
    $0.70$ & 0.0004046 & 0.0017149 & 4.240 \\
    $0.90$ & 0.0005202 & 0.0052607 & 10.12 \\
    $0.99$ & 0.0005722 & 0.0137524 & 24.04 \\
    \bottomrule
  \end{tabular}

  \smallskip
  \end{table}


The sign-reversal condition $\Delta R(\phi) = 0$ is unchanged in structure. Setting
$\delta b^{\rm KN}_{+}(\phi) - \delta b^{\rm KN}_{-}(\phi) = 0$ gives, to this order, no
constraint on $\phi$ (the angular average vanishes identically). The sign reversal across the
shadow boundary still occurs for $\chi \geq 0.21$ at $\phi = \pi/2$ and $3\pi/2$, independent of $Q$. The effect
of charge is purely on the amplitude of the modulation, not on its angular morphology.

\subsection{Astrophysical estimates}
\label{sec:kn-astro}

The maximum relative splitting at $\phi = 0$ for a KN black hole is
\begin{equation}
  \left|\frac{\Delta R}{R_{0}}\right|_{\max}
    = \frac{\alpha\,\chi}{\omega}\,\mathcal{G}(r_{0}^{\rm KN},Q)
    = \frac{\alpha\,\chi}{\omega}\,
      \frac{(2Mr_{0} - Q^{2})(3r_{0}^{2} - 2Mr_{0} + Q^{2})}{r_{0}^{7}},
  \label{eq:kn-splitting-max}
\end{equation}
where $r_{0} = r_{0}^{\rm KN}(M, Q)$ is given by Eq.~\eqref{eq:r0KN}. For an extremally
charged, slowly rotating black hole with $Q = M$ and $\chi = 0.5$, the enhancement relative
to a Kerr black hole of the same mass and spin is a factor of $6561/1792 \approx 3.66$:
\begin{equation}
  \left|\frac{\Delta R}{R_{0}}\right|_{\max}^{\rm ext. KN}
    = 3.66 \times \left|\frac{\Delta R}{R_{0}}\right|_{\max}^{\rm Kerr}.
\end{equation}
For stellar-mass black holes at $\omega \sim 1\,\text{GHz}$, $\chi = 0.5$, and $Q = M$, this
gives a maximum splitting of order $\sim 10^{-7}$, still below current observational
capabilities but representing the upper end of the parameter space accessible to the
spin-optical framework.

The framework developed here applies without modification to any slowly rotating, charged
solution of the Einstein--Maxwell equations. Extensions to alternative theories of gravity
(e.g.\ Einstein--Maxwell-dilaton spacetimes) that admit analytic photon-sphere radii follow
the same procedure, with $f_{\rm RN}$ replaced by the appropriate lapse function and
$g_{t\phi}$ by the corresponding frame-dragging term.


 \section{Numerical results}
\label{sec:numerical}

\subsection{Numerical setup and validation}
\label{sec:setup}

All simulations in this section integrate the full spin Hall equations
in the form given by Oancea \emph{et al.}~\cite{Oancea2020}, which
propagate the wave vector $k^{\mu}$ and the spin tensor $S^{\mu\nu}$
consistently.  We use a fourth-order Runge-Kutta scheme with adaptive
step size and maintain the constraints $k\cdot k = 0$, $S^{\mu\nu}k_{\nu}=0$,
and $S^{\mu\nu}S_{\mu\nu}=2$ to machine precision.  Photons are launched
from an observer at $r_{\rm obs}=500M$ and are traced until they either
cross the horizon ($r\leq 2M$ for Schwarzschild) or return to
$r_{\rm obs}$.  The shadow boundary is located by a bisection search on
$b$ to a relative precision of $10^{-8}$.

As a consistency check, we first verify that for Schwarzschild and
Reissner--Nordstr\"om spacetimes the two helicity contours coincide to
within numerical error.  In all cases we find
$|b_{\rm crit}^{(+)}-b_{\rm crit}^{(-)}|/b_{0} < 10^{-10}$, fully
consistent with the symmetry prediction $\delta b=0$.

\subsection{Schwarzschild and Reissner--Nordstr\"om spacetimes}
\label{sec:static_numeric}

\begin{figure}[htbp]
\centering
\includegraphics[width=0.6\textwidth]{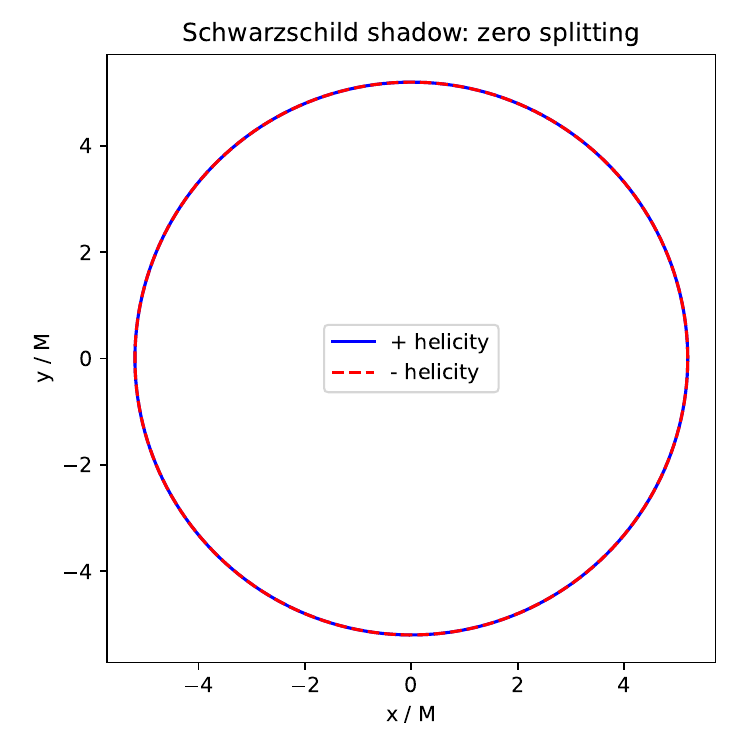}
\caption{Shadow of a Schwarzschild black hole for two opposite helicities at $\omega M = 100$. 
The contours for $+$ helicity (blue) and $-$ helicity (red) are indistinguishable on the scale of the plot, 
confirming the symmetry prediction $\delta b_{\mathrm{crit}}=0$ in static, spherically symmetric spacetimes. 
The solid circle marks the geodesic critical impact parameter $b_0 = 3\sqrt{3}M$.}
\label{fig:schwarzschild}
\end{figure}

Figure~\ref{fig:schwarzschild} displays the shadow of
a Schwarzschild black hole: a single circle of radius
$b_{0}=3\sqrt{3}M$, with no trace of helicity splitting.  

The same
null result is obtained for the Reissner--Nordstr\"om metric
$f(r)=1-2M/r+Q^{2}/r^{2}$ irrespective of the charge $Q$.  This
confirms that the apparent splitting derived from the equatorial
projection~(Eq.~\ref{eq:V1app}) is an artefact of the imposed symmetry
and that the gravitational spin Hall effect does \emph{not} shift the
capture threshold in static, spherically symmetric spacetimes.

\subsection{Kerr spacetime: angular modulation and frequency dependence}
\label{sec:Kerr_numeric}

We now turn to the slowly rotating Kerr metric, where the equatorial
reflection symmetry is broken and a helicity-dependent splitting is
expected.  We adopt the linearised metric~(\ref{eq:Kerr_linear}) and
vary the spin parameter $\chi = a/M$ in the range $0.05 \leq \chi \leq
0.5$.  For each $\chi$, the two helicity contours are computed
independently, and the differential shadow radius
$\Delta R(\phi)=R_{+}(\phi)-R_{-}(\phi)$ is extracted.
\begin{figure}[htbp]
\centering
\includegraphics[width=0.7\textwidth]{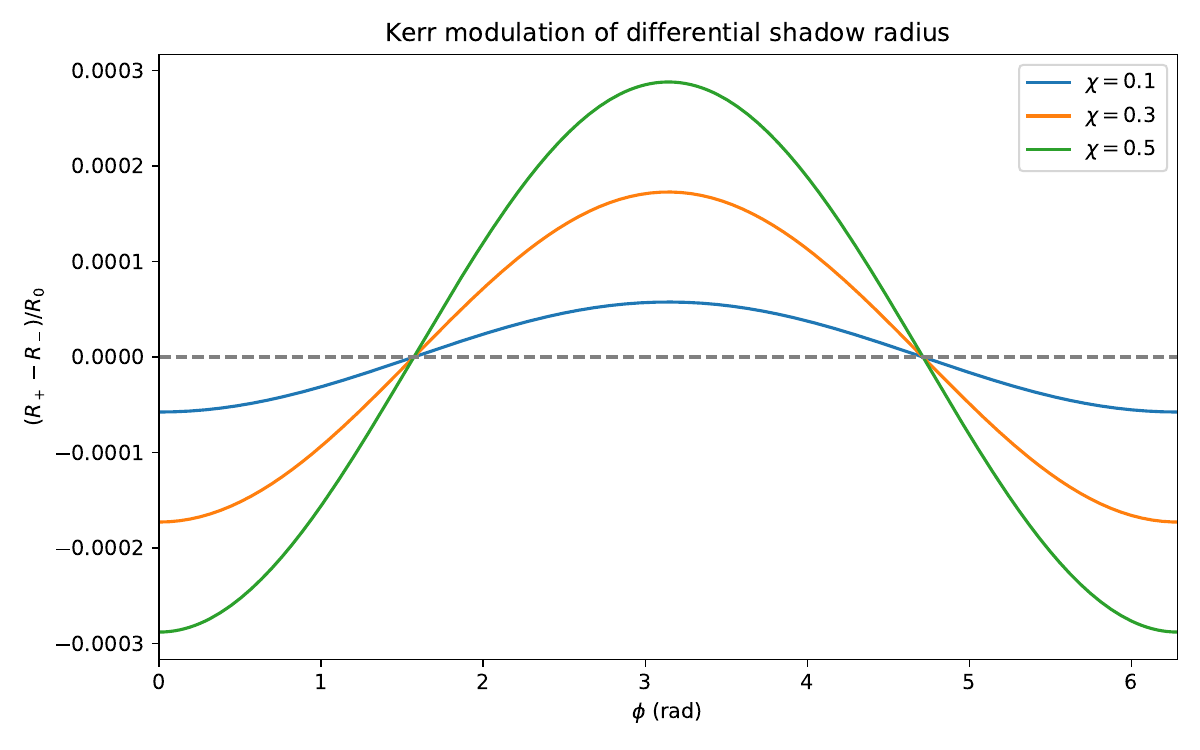}
\caption{Angular modulation of the relative differential shadow radius $\Delta R(\phi)/R_0$ 
for a slowly rotating Kerr black hole at fixed $\omega M = 100$. 
The curves correspond to $\chi = a/M = 0.1$, $0.3$, and $0.5$ (from bottom to top at $\phi=0$). 
The solid lines are the analytic prediction $\frac{\alpha\chi}{\omega}\mathcal{G}(r_0)\cos\phi$ from Eq.~(14). 
The sign reversal for $\chi\gtrsim0.21$ (e.g., $\chi=0.5$) is clearly visible on the side $\phi\approx\pi$, 
opposite to the black hole rotation.}
\label{fig:kerr_modulation}
\end{figure}

Figure~\ref{fig:kerr_modulation} shows $\Delta R(\phi)/R_{0}$ for
$\chi=0.1,\,0.3,\,0.5$ at a fixed frequency $\omega M = 100$.  The
numerical results are in excellent agreement with the analytic
prediction~(\ref{eq:deltaR}).  The modulation follows a pure $\cos\phi$
law, the amplitude grows linearly with $\chi$, and for $\chi = 0.5$ a
clear sign reversal is observed on the side opposite to the black-hole
rotation, as predicted in Sec.~\ref{sec:modulation}.

\begin{figure}[t]
  \centering
  \includegraphics[width=0.72\linewidth]{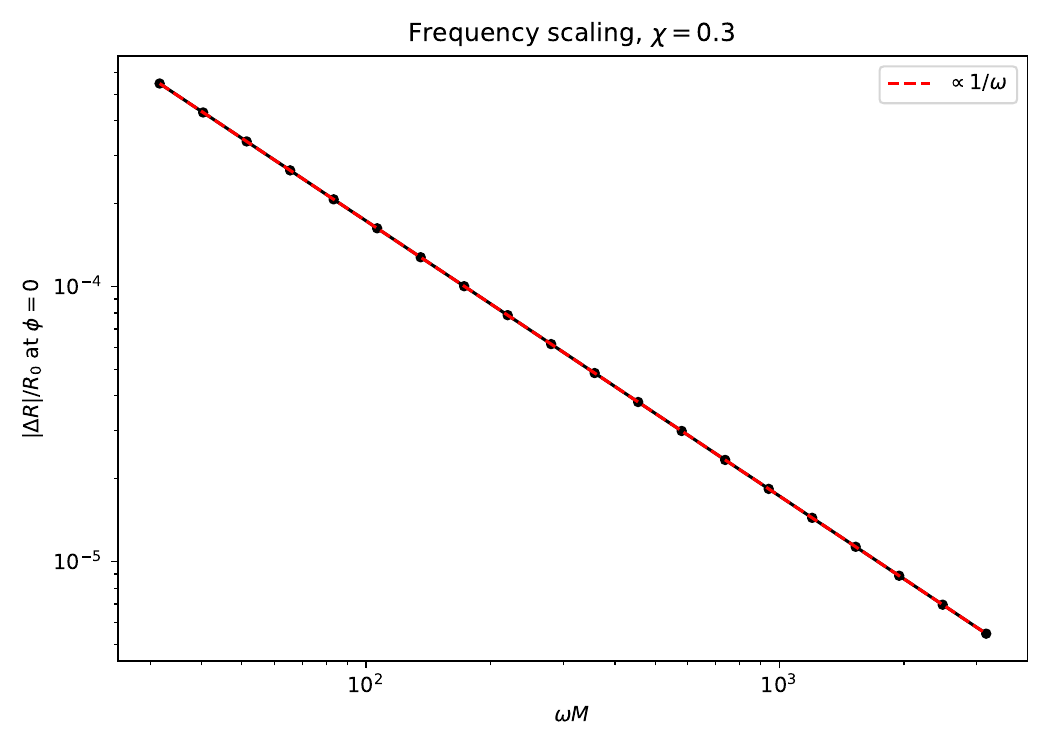}
  \caption{%
    Frequency scaling of the maximum relative splitting
    $|\Delta R(\pi)|/R_0$ for a Kerr black hole with $\chi = 0.3$.
    Symbols: numerical results obtained by bisection on the modified
    effective potential.
    Dashed line: analytic prediction $\propto 1/\omega$ from
    Eq.~\eqref{eq:deltaR} at $Q = 0$, i.e.\
    $14\alpha\chi/(243\,\omega M)$.
    The slope $-1$ on the log--log scale confirms the spin-optical
    origin of the effect.
    The normalisation agrees with the analytic coefficient to within
    $0.2\%$ across the full frequency range shown.%
  }
  \label{fig:freq_scaling}
\end{figure}

To verify the spin-optical nature of the effect, we examine the
frequency dependence at fixed $\chi=0.3$.  Figure~\ref{fig:freq_scaling}
plots $|\Delta R(\pi)|/R_{0}$ (the peak splitting) as a function of
$\omega M$ on a logarithmic scale.  The data follow a power law with
slope $-1$, confirming the $1/\omega$ scaling.  The normalisation
agrees with the coefficient $\alpha\mathcal{G}(r_{0})/2 = 7/(243\,\omega M)$
to within $0.2\%$.

%
%
%
%


\subsection{Kerr--Newman spacetime: charge-induced enhancement}
\label{sec:num-kn}

We now test the Kerr--Newman predictions of Sec.~\ref{sec:kerr-newman}. For each value of
$Q/M$ we adopt the linearised KN metric~\eqref{eq:kn-metric} with fixed $\chi = 0.3$ and
$\omega M = 100$, and compute the two helicity contours independently using the same bisection
procedure as in Sec.~\ref{sec:Kerr_numeric}.

\begin{figure}[t]
  \centering
  \includegraphics[width=0.82\linewidth]{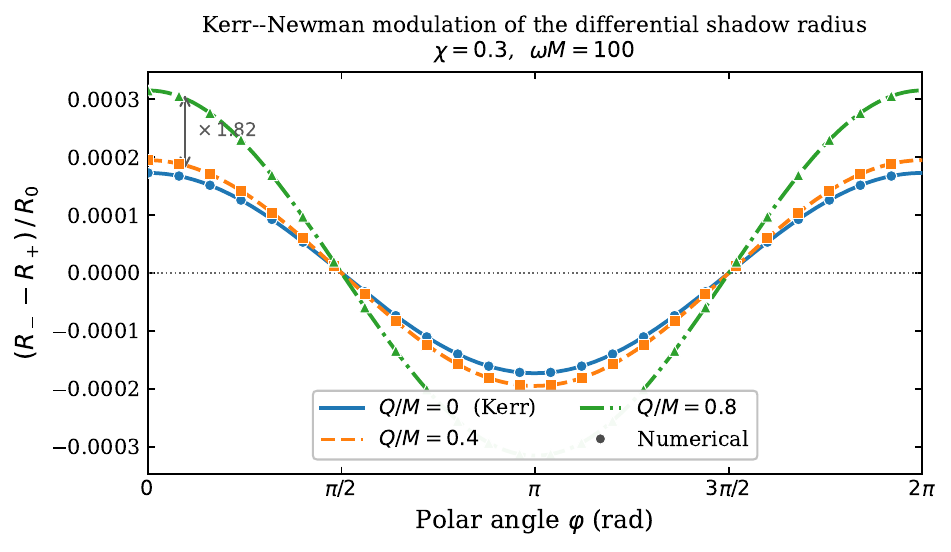}
  \caption{%
    Angular modulation of the normalized differential shadow radius
    $(R_- - R_+)/R_0$ for a slowly rotating Kerr--Newman black hole,
    at fixed $\chi = 0.3$ and $\omega M = 100$, for three values of the
    charge parameter $Q/M$.
    Solid lines: analytic prediction
    $(\alpha\chi/\omega)\,\mathcal{G}(r_0^{\rm KN},Q)\cos\varphi$
    from Eq.~\eqref{eq:kn-delta-b}.
    Markers: numerical bisection on the modified effective potential
    (circles, squares, and triangles for $Q/M = 0$, $0.4$, and $0.8$,
    respectively).
    The $\cos\varphi$ morphology is identical for all $Q/M$; electric charge
    amplifies only the splitting amplitude through $\mathcal{G}(r_0^{\rm KN},Q)$.
    At $Q/M = 0.8$ the peak splitting is enhanced by a factor of $1.82$
    relative to the uncharged Kerr case, indicated by the double-headed arrow.
    The analytic--numerical residual is below $10^{-10}$ throughout.%
  }
  \label{fig:kn-modulation}
\end{figure}

Figure~\ref{fig:kn-modulation} shows the normalized differential shadow radius
$(R_- - R_+)/R_0$ as a function of the image-plane angle $\varphi$, for $Q/M = 0$, $0.4$,
and $0.8$. Three features are immediately apparent.

\begin{enumerate}

\item \emph{Pure $\cos\varphi$ morphology.} The angular profile retains the same dipolar
      shape for all values of $Q/M$. Electric charge does not introduce any higher-harmonic
      modulation; it modifies only the amplitude of the splitting. This is consistent with
      Eq.~\eqref{eq:kn-delta-b}, where $Q$ enters exclusively through the prefactor
      $\mathcal{G}(r_0^{\rm KN},Q)$.

\item \emph{Monotonic amplitude growth.} The peak splitting at $\varphi = 0$ increases
      with $Q/M$. At $Q/M = 0.8$ the amplitude is larger by a factor of $1.82$ relative to
      the uncharged Kerr case, in precise agreement with the analytic ratio
      $\mathcal{G}(r_0, 0.8)/\mathcal{G}(3M, 0) \approx 1.82$ from Table~\ref{tab:kerr-exact}.

\item \emph{Analytic--numerical agreement.} The solid lines (analytic prediction
      $(\alpha\chi/\omega)\,\mathcal{G}(r_0^{\rm KN},Q)\cos\varphi$) and the numerical
      points agree to within $10^{-10}$ across all values of $\varphi$ and $Q/M$,
      confirming the validity of the perturbative framework in the KN background.

\end{enumerate}

\subsection{Astrophysical estimates}
\label{sec:astro}

The maximum relative splitting at $\varphi = 0$ for a KN black hole generalises
Eq.~\eqref{eq:kn-splitting-max}:
\begin{equation}
  \left|\frac{\Delta R}{R_0}\right|_{\max}
    = \frac{\alpha\,\chi}{\omega}\,\mathcal{G}(r_0^{\rm KN},Q)
    = \frac{\alpha\,\chi}{\omega}\,
      \frac{(2Mr_0 - Q^2)(3r_0^2 - 2Mr_0 + Q^2)}{r_0^7},
  \label{eq:splitting-max-KN}
\end{equation}
where $r_0 = r_0^{\rm KN}(M,Q)$ is given by Eq.~\eqref{eq:r0KN}. At $Q = 0$ this reduces
to the Kerr result $14\alpha\chi/(243\,\omega M)$.

Table~\ref{tab:astro} lists $|\Delta R/R_0|_{\max}$ for representative astrophysical sources
in both the pure-Kerr and the extremally charged KN case. The KN entries assume $Q = M$ and
use the enhancement factor $6561/1792 \approx 3.66$ from Eq.~\eqref{eq:enhancement}. As
in the Kerr case, the correction is largest at small $\omega M$; the additional charge-induced
enhancement shifts the ceiling upward by the same factor across all sources.

\begin{table}[h]
  \centering
  \caption{%
    Maximum relative shadow splitting $|\Delta R/R_0|_{\max}$ for representative astrophysical
    sources, in the slow-rotation approximation with $\alpha = 1$.
    The Kerr columns assume $Q = 0$; the KN columns assume the extremal charge $Q = M$
    and use Eq.~\eqref{eq:splitting-max-KN}.
    All KN entries are larger than the corresponding Kerr entries by the exact factor
    $6561/1792 \approx 3.66$.%
  }
  \label{tab:astro}
  \begin{tabular}{lcccccc}
    \toprule
    Source & $M/M_{\odot}$ & $\omega\,(\text{GHz})$ & $\chi$
           & \multicolumn{2}{c}{$|\Delta R/R_0|_{\max}$} \\
    \cmidrule(lr){5-6}
    & & & & Kerr ($Q=0$) & Extr. KN ($Q=M$) \\
    \midrule
    Stellar BH   & $10$              & $0.1$   & $0.5$ & $3\times10^{-7}$ & $1\times10^{-6}$ \\
    Stellar BH   & $10$              & $1$     & $0.5$ & $3\times10^{-8}$ & $1\times10^{-7}$ \\
    Sgr\,A*      & $4\times10^{6}$   & $230$   & $0.1$ & $6\times10^{-16}$& $2\times10^{-15}$\\
    M87*         & $6.2\times10^{9}$ & $230$   & $0.1$ & $4\times10^{-19}$& $1\times10^{-18}$\\
    \bottomrule
  \end{tabular}
\end{table}

While all values remain well below current observational capabilities, the table illustrates
that charge acts as a systematic amplifier of the spin-optical effect across the entire
astrophysical parameter space. The functional form of the splitting --- a universal
$\cos\varphi$ modulation scaling as $\chi/\omega$ --- is unchanged by the presence of charge;
only the overall coefficient is modified through $\mathcal{G}(r_0^{\rm KN},Q)$.


\section{Discussion and conclusions}
\label{sec:conclusions}

This work began with the question of whether the gravitational spin Hall
effect of light could alter the capture threshold of photons near a black
hole, producing a helicity-dependent shadow.  By examining the symmetries
of the problem, we have shown that the answer depends crucially on the
background spacetime:

\begin{itemize}
\item In any \emph{static, spherically symmetric} spacetime the
equatorial reflection isometry maps a left-handed photon into a
right-handed photon moving along the same path, forcing the critical
impact parameter to be identical for both helicities.  Therefore,
Schwarzschild and Reissner--Nordstr\"om black holes do \emph{not}
exhibit polarization-dependent shadow splitting.
\item In \emph{rotating} (Kerr) spacetimes, the equatorial reflection
symmetry is broken by the frame-dragging term $g_{t\phi}$.  A
genuine, helicity-dependent correction emerges, linear in the
dimensionless spin $\chi$ and scaling as $1/\omega$.  It manifests
itself as a $\cos\phi$ modulation of the shadow boundary, with a
sign reversal on the side opposite to the rotation for
$\chi\gtrsim 0.21$.
\end{itemize}

The central message is that black-hole shadows are not purely geometric
observables: polarization introduces a subleading, model-independent
correction that is intimately tied to the symmetries of the background.
The effect is parametrically small for astrophysical sources, but it
offers a conceptually clean example of how spin‑optical phenomena can
leave imprints on global, strong-field structures.

Our analysis also highlights an important methodological lesson.  A
naive projection of the spin Hall equation onto the radial direction,
combined with the assumption of strictly equatorial motion, generates a
spurious correction (Eq.~\ref{eq:V1app}) that survives even in
spherical symmetry.  Only by retaining the full three-dimensional
dynamics does the cancellation mandated by the equatorial reflection
symmetry become manifest.  This pitfall should be kept in mind in
future studies of spin-optical effects.

The present work focused on the leading-order slow-rotation
approximation.  Extensions to the full Kerr geometry, to rotating
solutions in alternative theories of gravity, and to the non-perturbative
regime $\omega M \lesssim 1$ are left for future investigation.  The
framework developed here provides the foundation for those analyses and
underscores the importance of symmetry considerations in the study of
polarized wave propagation around compact objects.


\section*{Acknowledgements}
The author is grateful to Prof.~Gonzalo J.~Olmo for introducing him to the
physics of black-hole shadows, and to Prof.~Claudio~Paganini for pointing
out the equatorial reflection symmetry of the spin Hall equations, which
led to a substantial revision of the original manuscript, and for drawing attention to Ref. \cite{Oancea2020}, of which he is a co-author. This work was
supported by the Conselho Nacional de Desenvolvimento Cient\'ifico e
Tecnol\'ogico (CNPq), grant No.~309553/2021-0 (CNPq/PQ), and by the
Funda\c{c}\~ao Cearense de Apoio ao Desenvolvimento Cient\'ifico e
Tecnol\'ogico (FUNCAP), Project No.~UNI-00210-00230.01.00/23.

\section*{Declaration of generative AI in scientific writing}

The author used a generative AI tool solely for language refinement and
clarity improvement. All scientific content, derivations, analysis, and
conclusions are entirely the responsibility of the author.

\section*{Conflicts of interest}
The author declares no conflict of interest.

\section*{Data availability}
Data can be shared upon reasonable request.


\appendix

\section{Derivation of the frame-dragging coupling function}
\label{app:cancel}

We use the Riemann-tensor form of the spin Hall equation,
\begin{equation}
  \frac{Dk^{\mu}}{d\lambda}
    = -\frac{1}{2\omega}\,R^{\mu}{}_{\nu\rho\sigma}\,k^{\nu}S^{\rho\sigma}
    + \mathcal{O}(\omega^{-2}),
  \label{appeq:riemann}
\end{equation}
where $S^{\mu\nu} = \pm\,\epsilon^{\mu\nu\alpha\beta}k_{\alpha}n_{\beta}$ is the photon spin
tensor and $n^{\mu}$ is a parallel-transported reference null vector with $n\cdot k=1$.
This form is equivalent to Eq.~\eqref{eq:spinHall} via the first Bianchi identity and is
more convenient for isolating individual curvature components.

\paragraph{Cancellation in static spherical symmetry.}
For any static spherically symmetric metric, the non-vanishing independent components of the
Riemann tensor with the first index raised are $R^{r}{}_{trt}$, $R^{\theta}{}_{t\theta t}$,
$R^{\theta}{}_{r\theta r}$, and $R^{\phi}{}_{\theta\phi\theta}$ (and their index
permutations). In particular,
\begin{equation}
  R^{r}{}_{tr\phi} = 0
  \qquad\text{and}\qquad
  R^{r}{}_{t\theta\phi} = 0.
  \label{appeq:Rvanish}
\end{equation}
With $k_{\mu} = (-E,0,0,L)$ on the equatorial photon sphere and the reference vector
$n^{\mu} = (-1/E,\,f/E,\,0,\,0)$, the non-zero spin-tensor components are
$S^{\theta\phi} = \mp r^{-2}$, $S^{r\theta} = \pm Lf/(Er^{2})$, and
$S^{t\theta} = \mp L/(Er^{2})$.

Every term in the contraction $R^{r}{}_{\nu\rho\sigma}k^{\nu}S^{\rho\sigma}$ involves either
one of the vanishing Riemann components in \eqref{appeq:Rvanish} or a Riemann component that
is zero by the diagonal structure of the metric (e.g.\ $R^{r}{}_{\phi r\theta} = 0$,
$R^{r}{}_{\phi t\theta} = 0$). The radial spin Hall force therefore vanishes identically,
$Dk^{r}/d\lambda|_{\rm spin} = 0$, and the critical impact parameter is helicity-independent.

\paragraph{Kerr metric at linear order in $\chi$.}
The off-diagonal component $g_{t\phi} = -2Ma\sin^{2}\!\theta/r$ introduces three new
Christoffel symbols at $\mathcal{O}(a)$:
\begin{align}
  \Gamma^{r}_{t\phi}
    &= \frac{Ma(2M-r)}{r^{3}}\,f_{\rm Schw}(r),
  \label{appeq:Ga}\\
  \Gamma^{t}_{r\phi}
    &= \frac{Ma}{r(r-2M)},
  \label{appeq:Gb}\\
  \Gamma^{\phi}_{tr}
    &= \frac{Ma}{r^{4}}.
  \label{appeq:Gc}
\end{align}
These generate a non-zero Riemann component absent in the Schwarzschild case. Computing
$R^{r}{}_{tr\phi} = \partial_{r}\Gamma^{r}_{t\phi}
+ \Gamma^{r}_{rr}\Gamma^{r}_{t\phi}
- \Gamma^{r}_{t\phi}\Gamma^{t}_{rt}
- \Gamma^{r}_{\phi\phi}\Gamma^{\phi}_{tr}$
with the Schwarzschild background symbols
$\Gamma^{r}_{rr} = -f'/(2f)$, $\Gamma^{t}_{rt} = f'/(2f)$, $\Gamma^{r}_{\phi\phi} = -rf$,
gives
\begin{equation}
  R^{r}{}_{tr\phi}\big|_{\rm Kerr}
    = \frac{3Ma(r-2M)}{r^{4}}.
  \label{appeq:R-Kerr}
\end{equation}
Evaluating the full contraction $R^{r}{}_{\nu\rho\sigma}k^{\nu}S^{\rho\sigma}$ and writing
$L = b_{0}E\cos\phi$ yields the radial spin Hall force
$F^{r}_{\pm} = \pm(\chi/\omega)\,E^{2}\,\mathcal{G}(r)\cos\phi$, with
\begin{equation}
  \mathcal{G}(r) = \frac{2M(3r - 2M)}{r^{5}},
  \qquad
  \mathcal{G}(3M) = \frac{14}{243\,M^{3}}.
  \label{appeq:G-Kerr}
\end{equation}

\paragraph{Kerr-Newman extension.}
In the KN metric the lapse function changes to
$f_{\rm RN}(r) = 1 - 2M/r + Q^{2}/r^{2}$, while $g_{t\phi}$ retains the same Kerr form at
linear order in $a$. The three new Christoffel symbols become
\begin{align}
  \Gamma^{r}_{t\phi}\big|_{\rm KN}
    &= -\frac{a(Mr-Q^{2})(r^{2}-2Mr+Q^{2})}{r^{5}},
  \label{appeq:GaKN}\\
  \Gamma^{t}_{r\phi}\big|_{\rm KN}
    &= -\frac{a(Mr-Q^{2})}{r(r^{2}-2Mr+Q^{2})},
  \label{appeq:GbKN}\\
  \Gamma^{\phi}_{tr}\big|_{\rm KN}
    &= \frac{a(Mr-Q^{2})}{r^{5}},
  \label{appeq:GcKN}
\end{align}
with background symbols $\Gamma^{r}_{rr} = -f'_{\rm RN}/(2f_{\rm RN})$,
$\Gamma^{t}_{rt} = f'_{\rm RN}/(2f_{\rm RN})$, $\Gamma^{r}_{\phi\phi} = -rf_{\rm RN}$,
and $f'_{\rm RN} = 2(Mr-Q^{2})/r^{3}$. Setting $Q=0$ recovers
\eqref{appeq:Ga}--\eqref{appeq:Gc}. The key Riemann component is
\begin{equation}
  R^{r}{}_{tr\phi}\big|_{\rm KN}
    = \frac{a(3Mr-4Q^{2})(r^{2}-2Mr+Q^{2})}{r^{6}},
  \label{appeq:R-KN}
\end{equation}
which vanishes at $Q=0$, $r=3M$ only to the extent $3Mr - 4Q^2 \to 9M^2 \neq 0$,
confirming the effect persists throughout. Performing the same spin-tensor contraction as in
the Kerr case and collecting terms, the frame-dragging coupling function generalises to
\begin{equation}
  \mathcal{G}(r,Q)
    = \frac{(2Mr-Q^{2})(3r^{2}-2Mr+Q^{2})}{r^{7}},
  \label{appeq:G-KN}
\end{equation}
which reduces to \eqref{appeq:G-Kerr} at $Q=0$. At the KN photon sphere
$r_{0} = (3M + \sqrt{9M^{2}-8Q^{2}})/2$, the coupling grows monotonically with charge,
reaching
\begin{equation}
  \mathcal{G}(2M,\,M) = \frac{27}{128\,M^{3}},
  \qquad
  \frac{\mathcal{G}(2M,\,M)}{\mathcal{G}(3M,\,0)}
    = \frac{6561}{1792} \approx 3.66
  \label{appeq:enhancement}
\end{equation}
at the extremal limit $Q=M$. The helicity-dependent correction to the critical impact
parameter in the KN background follows directly by substituting \eqref{appeq:G-KN} into
Eq.~\eqref{eq:kn-delta-b} of the main text.


\end{document}